\sfcode`A=1000 \sfcode`B=1000 \sfcode`C=1000 \sfcode`D=1000
\sfcode`E=1000 \sfcode`F=1000 \sfcode`G=1000 \sfcode`H=1000
\sfcode`I=1000 \sfcode`J=1000 \sfcode`K=1000 \sfcode`L=1000
\sfcode`M=1000 \sfcode`N=1000 \sfcode`O=1000 \sfcode`P=1000
\sfcode`Q=1000 \sfcode`R=1000 \sfcode`S=1000 \sfcode`T=1000
\sfcode`U=1000 \sfcode`V=1000 \sfcode`W=1000 \sfcode`X=1000
\sfcode`Y=1000 \sfcode`Z=1000
\documentstyle[12pt,aasms4] {article}

\begin{document}

\title{Infrared to Ultraviolet Wavelength-Dependent Variations Within the Pulse Profile Peaks of the Crab Nebula Pulsar}

\author{S.S. Eikenberry, G.G. Fazio, S.M. Ransom}
\affil{Harvard-Smithsonian Center for Astrophysics, Cambridge, MA 02138}

\author{J. Middleditch}
\affil{Los Alamos National Laboratory, Los Alamos, NM 87545}

\author{J.A. Kristian}
\affil{Observatories of the Carnegie Institution of Washington, Pasadena, CA 91101}

\author{C.R. Pennypacker}
\affil{Space Sciences Laboratory, University of California, Berkeley, CA 94720}

\begin{abstract}

	We present evidence of wavelength-dependent variations within
the infrared, optical, and ultraviolet pulse profile peaks of the Crab
Nebula pulsar.  The leading and trailing edge half-width half-maxima
of the peaks display clear differences in their wavelength
dependences.  In addition, phase-resolved infrared-to-ultraviolet
color spectra show significant variations from the leading to trailing
edges of the peaks.  The color variations between the leading and
trailing edges remain significant over phase differences smaller than
0.0054, corresponding to timescales of $<180 \mu$s.  These results are
not predicted by any current models of the pulsar emission mechanism
and offer new challenges for the development of such models.

\end{abstract}


\section{Introduction}

	Astronomers have been studying pulsars for nearly 30 years and
yet, we still have no clear understanding of the emission mechanisms
that produce the trademark pulsations.  However, pulsar emission
models do exist for some pulsars, in particular the $\gamma$-ray
pulsars, of which the Crab Nebula pulsar is the most well-studied.  In
current models of Crab Nebula pulsar emissions, primary energy
generation occurs in particle accelerator ``gaps'' in the outer
magnetosphere (i.e. Cheng, Ho, and Ruderman 1986ab; Chiang and Romani
(1994); Romani and Yadigaroglu (1995) - hereafter CHRab, CR94, and RY,
respectively).  The accelerated particles produce $\gamma$-rays
through curvature radiation, and these $\gamma$-rays interact with the
magnetosphere in a complex manner to produce the X-ray, UV, optical,
and infrared pulsations, so that the non-radio emissions are closely
linked to each other.  Because of this linkage, we are investigating
the pulsar emission mechanism through the UV, optical, and infrared
emissions, where high signal-to-noise observations are more practical
to obtain than in the X-ray and $\gamma$-ray wavebands (see Ransom
{\it et al.} (1994) for earlier results).

	While the emission models mentioned above vary in their
underlying assumptions, they share several common features.  Not least
among these is the assumption that the emission arises in an
essentially uniform region, with the observed shape and sharpness of
the pulse profile peaks\footnote{We refer to the 2 major features in
the profile as Peak 1 and Peak 2, joined by the Bridge.  This is to
avoid confusion, as some previous authors refer to the Bridge as the
``Interpulse'', while others use that name for Peak 2.} (see Figure 1)
resulting from a combination of time-of-flight delays and relativistic
aberration of the magnetic field lines.  In some of these models
(CHRab), the mapping from observed pulse phase to position in the
emission region is neither one-to-one nor continuous, so that the
observed flux in a given pulse phase interval is a combination of
emission from multiple disjoint sections of the emission region, and
the locations of the particular sections which contribute emission to
that phase interval depend on the observer's viewing angle.  This
mixing of physical regions within the observed phase interval
effectively averages the emission over a range within the emission
region, with the largest amount of averaging at the peak maximum.
Thus, the observed emission properties should remain constant or
change slowly and smoothly over the peak, and should certainly not
exhibit any sharp changes near the peak maximum.  It is with this
prediction in mind that we examine the peaks of the Crab Nebula pulsar
for wavelength-dependent variations about the peak maximum.

\section{Observations and Data Reduction}

	We made the infrared observations for this work in the J
($1.25 \mu$m), H ($1.65 \mu$m), and K ($2.2 \mu$m) bands using the
Solid-State Photomultiplier (SSPM) instrument on the Multiple Mirror
Telescope\footnote{The MMT is jointly owned and operated by the
University of Arizona and the Smithsonian Astrophysical Observatory.}
on January 18-22, 1995.  The SSPM instrument (Eikenberry, {\it et
al.}, 1996) is a high-speed near-infrared photometer, with
single-photon-counting performance and submicrosecond time resolution,
based on the Solid-State Photomultiplier detector developed by the
Rockwell International Science Center (Petroff, {\it et al.}, 1987).
For these observations, we recorded counts from the SSPM using $20
\mu$s time bins, with an EG\&G rubidium frequency standard for timing
reference.  The resulting time-series were corrected to the solar
system barycenter and used to determine the pulsar timing
characteristics (see Eikenberry, {\it et al.} (1996) for details).  We
determine the pulsar frequency to be $f_0 = 29.9047918 \pm 0.0000013$
Hz and the frequency derivative to be $\dot f = -3.83 \pm 0.12 \times
10^{-10}$ Hz/s at MJD 49736.17666605, in good agreement with the
Jodrell Bank timing ephemeris (A.G. Lyne, personal communication).
Given this timing solution, we then create pulse profiles for the Crab
Nebula pulsar in each of the J,H, and K wavebands by folding the
barycentered time-series at the appropriate values of $f$ and $\dot
f$, and then background-subtracting the profiles.  In order to augment
these data, we also include optical (V) and UV profiles taken with the
HST High-Speed Photometer (HSP) (Percival {\it et al.}, 1993) in the
analyses.

\section{Analysis}

	In order to study wavelength-dependent variations in the peaks
of the Crab Nebula pulsar, we divide each peak into 2 halves - the
leading and trailing edges - with the emission maximum marking the
point of division.  We illustrate the resulting phase conventions for
the V-band in Figure 1.  As mentioned above, we expect that the
emission properties should not change sharply near the emission
maximum.  Therefore, the first step in the analysis is to examine the
wavelength dependences of the shape about the peak maximum - in this
case the half-width half-maxima of the leading and trailing edges of
the peaks.  Next, we analyze the color spectra of the peaks for
significant variations between the leading and trailing edges.
Finally, we perform analyses to reveal the range of ``timescales'' for
the phase-resolved color variations in Peak 1.

\subsection{Peak half-width half-maximum} \label{hwhm}

	We measure the half-width half-maximum (HWHM) by simply taking
the difference in phase between the peak maximum and the points where
the flux drops to 1/2 of the maximum value.  While this procedure is
straightforward, it does not easily allow us to determine the
uncertainties in the HWHM values due to the Poisson statistics of the
pulse profiles.  In order to estimate these uncertainties, we use a
Monte Carlo simulation, proceeding as follows.  First, we fit independent
third-order polynomials to the leading and trailing edges of the
peaks, and measure the HWHM from the points where the fits equal
one-half of the peak maximum.  (In all cases, the fit determination of
the HWHM matches the manually measured HWHM.)  Next, we take the phase
bins in the peak region and assume that the noise in the number of
counts in each bin follows a Poisson distribution.  We then take the
fit to the peak and add to each bin a normally-distributed random
number with a variance corresponding to the Poisson noise for that
bin.  We fit this new profile and measure the new half-widths, and
then repeat the procedure 1000 times for each peak.  Finally, we take
the standard deviation of the simulated half-widths to be the $1
\sigma$ uncertainty in the measured value.  The resulting HWHM values
and uncertainties for Peak 1 are plotted in Figure 2.

	The result of this analysis is that the wavelength dependences
of the HWHM for the leading and trailing edges are noticeably
different.  The Peak 1 leading edge HWHM rises from the UV to J-band,
and then shows a break at $1.25 \mu$m, declining from J-band to
K-band.  Meanwhile, the Peak 1 trailing edge HWHM rises slowly from UV
to J-band or H-band, and then sharply to K-band.  While the HWHM
values for Peak 2 are noisier, if we divide the Peak 2 leading edge
HWHM by the trailing edge HWHM and fit the resulting ratio with a flat
line, we find $\chi ^2 = 61$ for 4 degrees of freedom.  This indicates
a significant difference between the wavelength dependences for the
leading and trailing edge HWHM for Peak 2.  Although previous work has
shown weak evidence of similar effects (Pravdo and Serlemitsos, 1981),
these data present the first significant evidence of
wavelength-dependent variations within the individual peaks of the
Crab Nebula pulsar.

\subsection{Phase-resolved spectra}

	We can also investigate the presence of wavelength-dependent
variations through the color spectra of the peaks.  The time
resolution of our SSPM data and the HST HSP data allows us to
construct phase-resolved spectra of the Crab Nebula pulsar.  We divide
the pulse profiles into phase segments, and calculate the fraction of
the total photon counts received in each segment using the phase
conventions shown in Figure 1.  We then take these fractional photon
fluxes from each waveband for a given phase segment to produce a
relative color ``spectrum''.  Note that the normalization between
bands in the relative color spectrum is essentially arbitrary.
However, these relative, unnormalized color spectra are the most
sensitive tools we can use to analyze color differences or
similarities between different phase intervals of the pulse profile,
given the large uncertainties introduced by reddening corrections to
the Crab Nebula pulsar.

	We present the relative color spectra for the leading and
trailing edges of Peak 1 in Figure 3(a).  Note that there is a
distinct difference between the spectral shapes of the leading and
trailing edges of the peak.  Similarly, in Figure 3(b), we present the
relative spectra for the leading and trailing edges of Peak 2.  Again,
the leading and trailing edges exhibit markedly different colors.
Thus, these results confirm the presence of wavelength-dependent
variations in the peaks of the Crab Nebula pulsar.

\subsection{Rapid color variability with phase}

	We present a series of unnormalized Peak 1 color spectra with
a phase resolution corresponding to 1 ms in Figure 4.  Note that the
spectral shape changes as a function of phase on the 1 ms timescale of
the frames.  Furthermore, the difference in spectral shape between the
Peak 1 leading and trailing edges (see Figure 3) appears clearly even
in the two 1-ms frames bracketing the peak maximum (phase = 0).

	In order to determine the minimum phase difference for which
this color variation is resolvable, we perform the following analysis.
First, for a given phase interval, we calculate the fractional photon
flux before and after the peak maximum in each waveband.  We then
divide the leading edge photon fluxes by the trailing edge photon
fluxes, resulting in values which should all be equal (within errors)
for the case of no leading/trailing color variation, and we calculate
$\chi ^2$ for the variation of this ratio from a flat line.  We must
also take into account the uncertainties in the peak location and do
so by repeating the procedure for all possible combinations of the
Peak 1 location in the 5 pulse profiles (within $1 \sigma$
uncertainties).  We then take the minimum $\chi ^2$ value from these
combinations as the limiting case for the significance of color
variability over the phase interval in question, and we repeat this
analysis over a range of phase intervals down to $20 \mu$s.  The
shortest phase interval which shows significant variations at the
99.9\% level is 0.0054, corresponding to a time of $180 \mu$s.  Thus
we conclude that the minimum ``timescale'' for color variations with
phase within Peak 1 is $< 180 \mu$s.

\section{Discussion}

	The above observations and analyses reveal significant
evidence of wavelength-dependent variations within the peaks of the
Crab Nebula pulsar's pulse profile.  The difference in the wavelength
dependence between the leading and trailing edge HWHM of Peak 1
(Figure 2) establishes the existence of such variations for the first
time.  In order to exhibit such behavior, the emission process must
somehow change around the peak maximum.  The phase-resolved color
spectra only strengthen this conclusion, exhibiting variability on
timescales from milliseconds to $<180 \mu$s, with clear differences
between the leading and trailing edges.

	The presence of wavelength-dependent variations within the
profile peaks presents some interesting questions regarding the origin
of the pulse shape.  As mentioned above, in current models the shape
of the Crab Nebula pulsar peaks is determined by time-of-flight delays
and relativistic aberration of the magnetic field lines.  The observed
sharp cusp of the peak maximum results from the simultaneous arrival
of flux from many (possibly disjoint) physical locations within the
emission region.  For those models in which the observed emission in
the peak maximum emanates from such disjoint physical regions (CHRab),
the emission properties should not change sharply near this phase
interval.  While it is possible that other models (CR94, RY95) may
allow such sharp variations near the maximum of Peak 1 in particular,
as the mapping there from the physical emission region to the observed
pulse phase may be more nearly monotonic, these models do not
currently predict such variations.

  	We can further quantify the characteristics of the variations
using the fact that the Peak 1 color spectrum varies on timescales at
least as rapidly as $180 \mu$s.  We believe both from models (CHRab,
CR94, RY95) and observations (Smith {\it et al.}, 1988) that the Crab
Nebula pulsar's high-energy emissions originate near the light
cylinder, where the magnetospheric corotation velocity approaches the
speed of light.  Thus, we conclude that the typical coherence length
for the emission region is similar to or less than $1.8 \times 10^{-4}
s \times c = 54$ km.

\section{Conclusions}

	We have presented evidence of wavelength-dependent variations
within the pulse profile peaks of the Crab Nebula pulsar.  This
evidence includes differences in the wavelength dependences of the
leading and trailing edge peak HWHM and color differences on
timescales from milliseconds to $<180 \mu$s.  This type of variation
is not predicted by current pulsar emission theories, and offers new
challenges for the development of such theories.

\acknowledgments

	We would like to thank K. Hays, M. Stapelbroek, and R.
Florence of the Rockwell International Science Center for providing
the SSPM detectors and invaluable advice and support; J. Dolan and the
HST HSP team for providing the optical and UV profiles; J. Geary, P.
Crawford, C.  Hughes of the CfA for help in constructing the SSPM
instrument; D.  Paolucci of the MIT Laser Spectroscopy Lab for help
with the laser alignment; W. Riley of EG\&G for providing the rubidium
frequency standard; R. Lucinio of Caltech for maintaining the Wizards;
the MMTO support crew for their assistance and donation of engineering
time; A.G. Lyne of Jodrell Bank for supplying the Jodrell Bank pulsar
ephemeris before publication; and R. Narayan, J.  Grindlay, and F.
Seward of the CfA and T.N.  Rengarajan of the Tata Institute for
Fundamental Research for their helpful discussions of this work.  S.
Eikenberry is supported by a NASA Graduate Student Researchers Program
grant through Ames Research Center.

\vfill \eject

\begin{figure} \vspace*{130mm}
\includegraphics{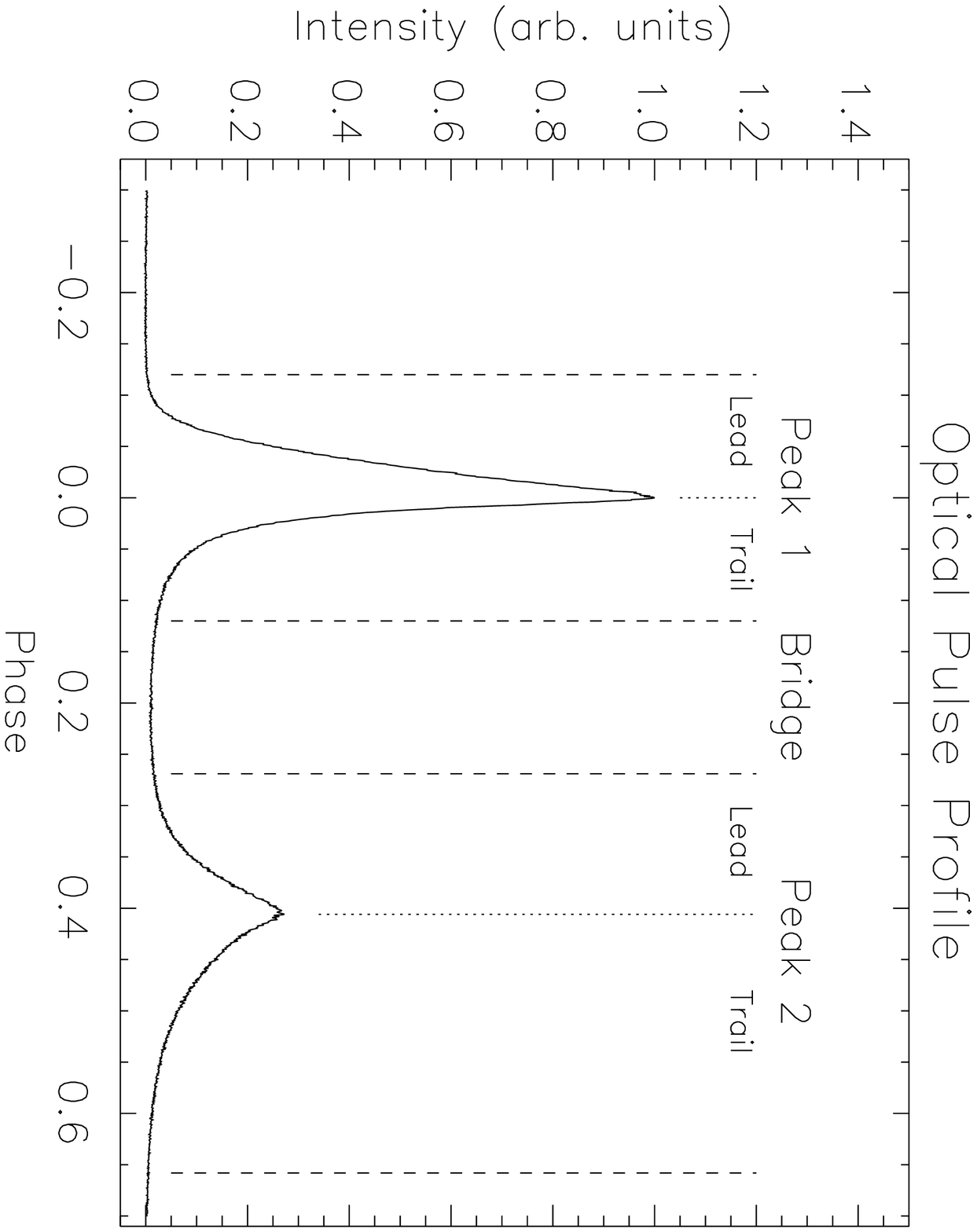} 
\caption{Pulse profile with phase conventions used in the analyses}
\end{figure}

\begin{figure} \vspace*{130mm}
\includegraphics{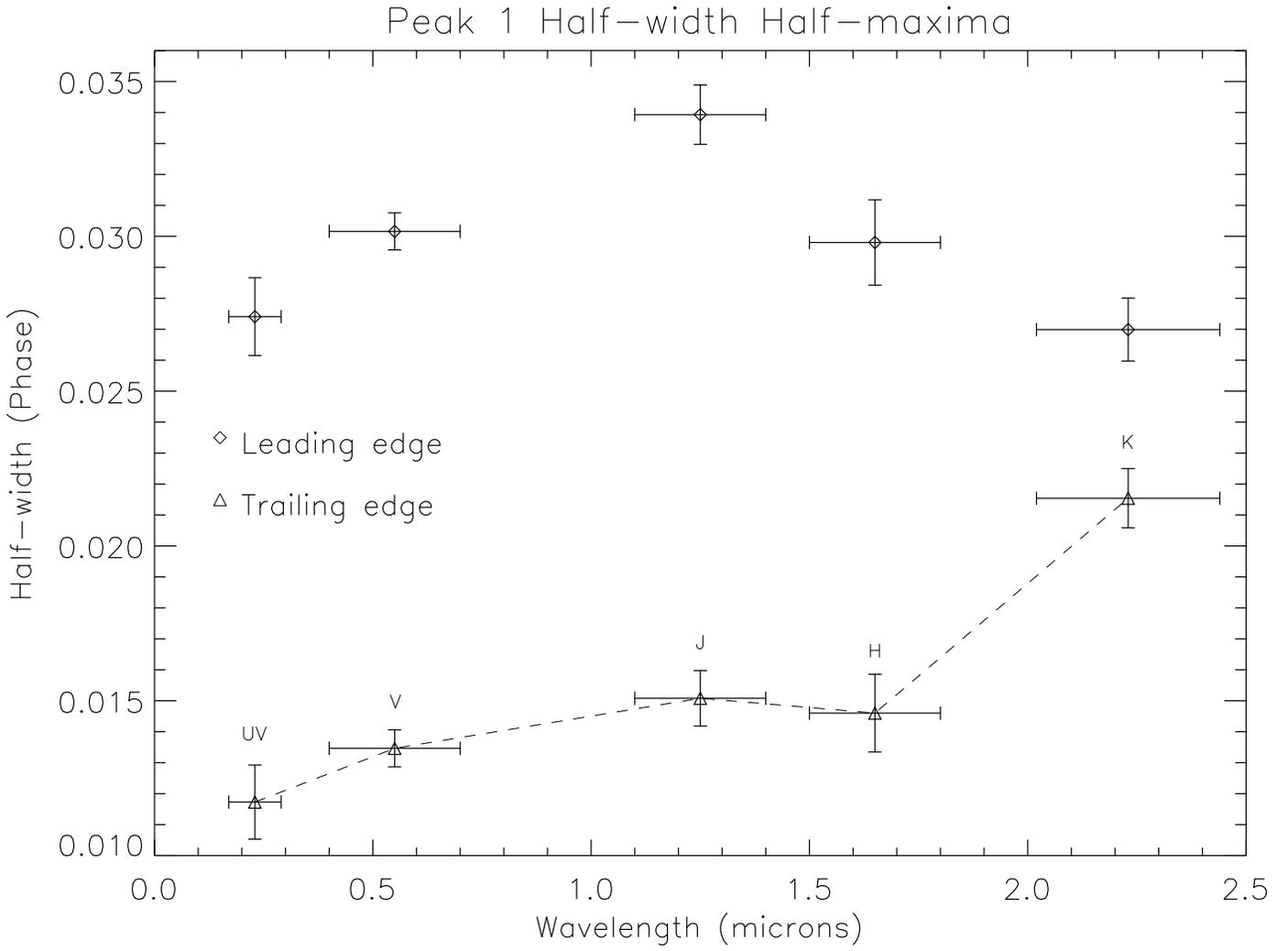} 
\caption{Peak half-width half-maximum versus wavelength for Peak 1 leading and trailing edges} 
\end{figure}

\begin{figure} \vspace*{130mm} 
\includegraphics{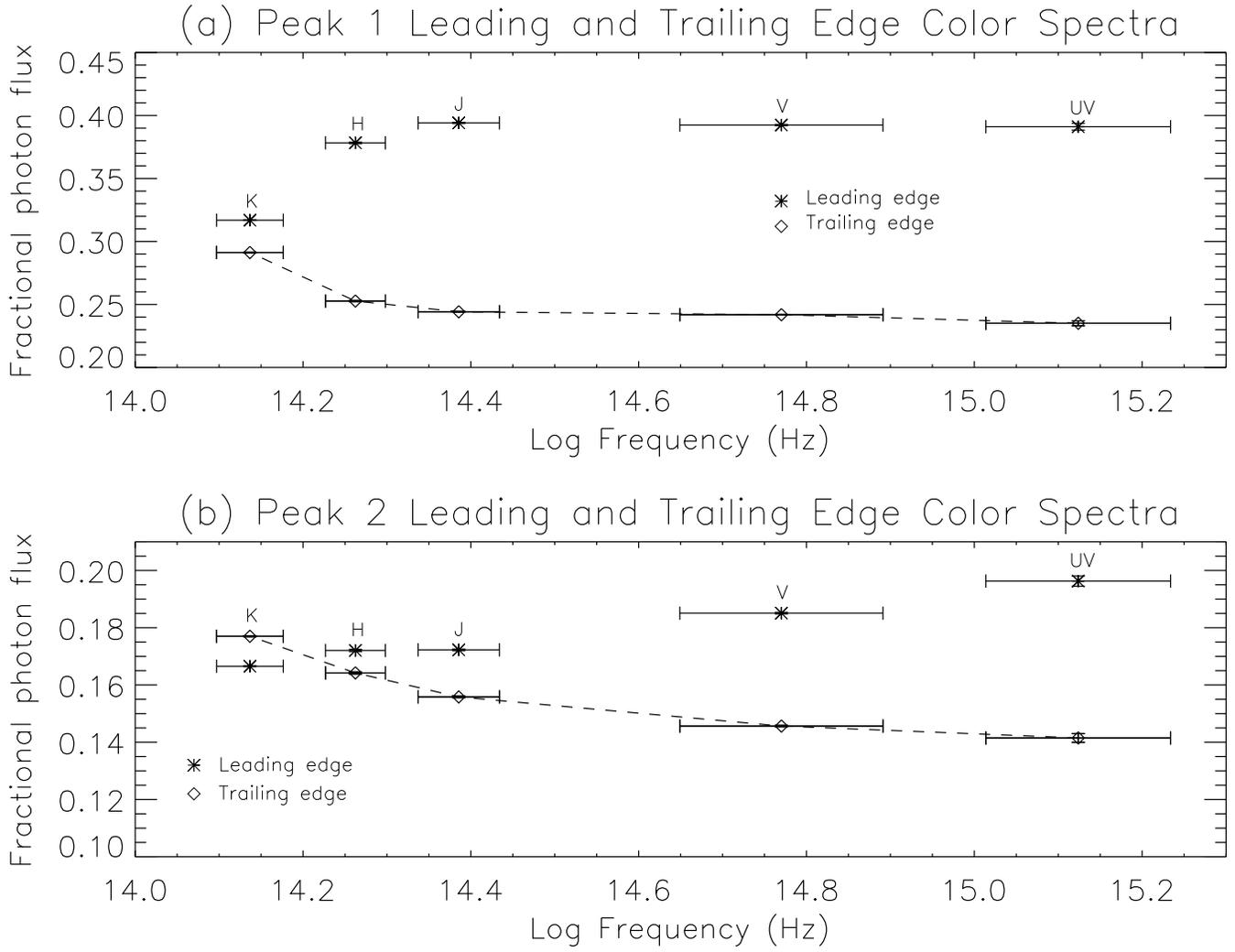} 
\caption{Fractional photon flux versus frequency for leading and trailing edges of (a) Peak 1, (b) Peak 2} 
\end{figure}

\begin{figure} \vspace*{130mm} 
\includegraphics{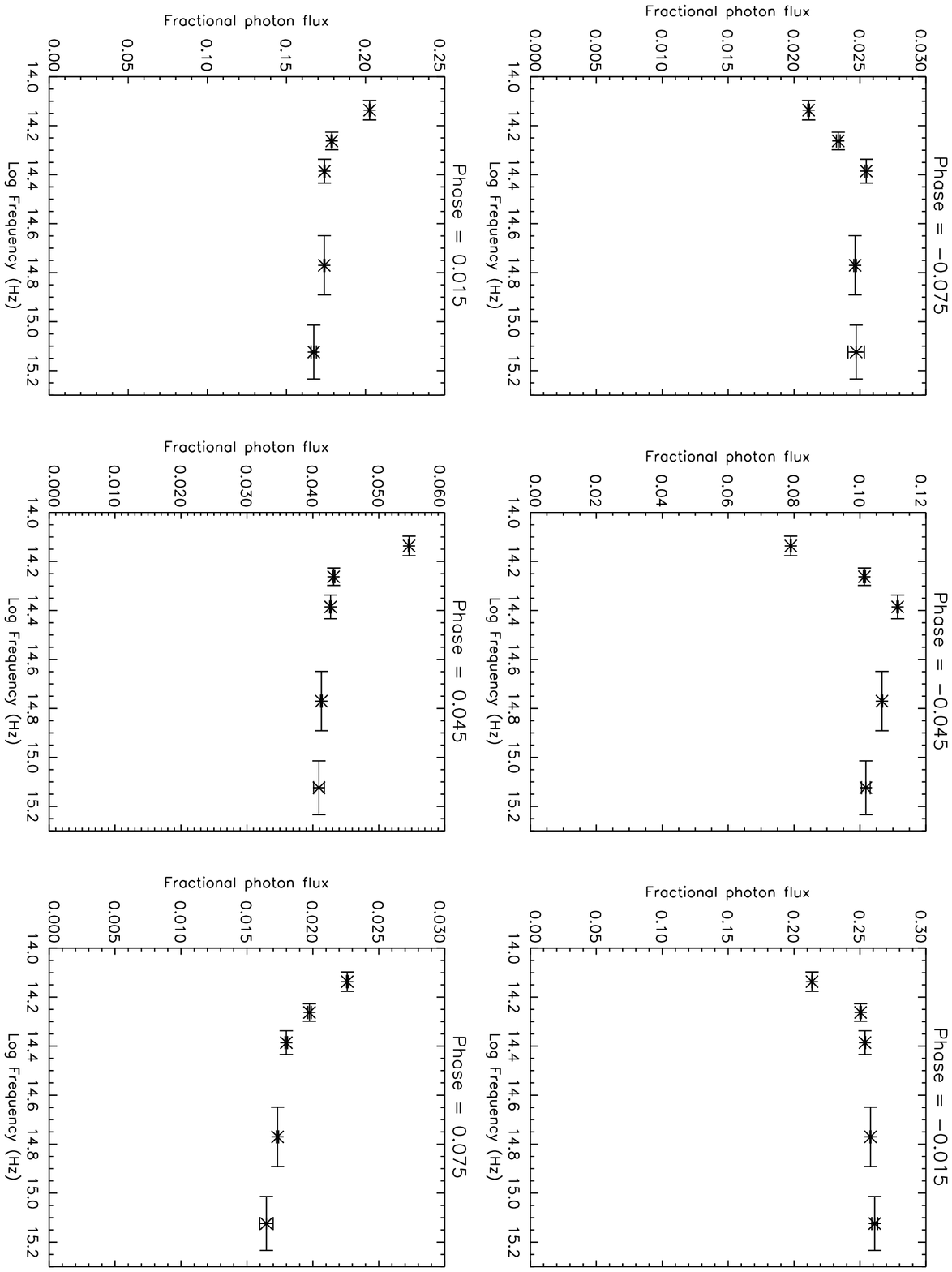} 
\caption{Series of phase-resolved, unnormalized Peak 1 color spectra.  The phase difference from frame to frame corresponds to $\sim 1$ ms.  Note the sharp change in spectral shape near phase=0.}
\end{figure}


\begin{references} 
\reference{CHRa} Cheng, K.S., Ho C., and Ruderman, M.
1986a, \apj, 300, 500 

\reference{CHRb} Cheng, K.S., Ho C., and Ruderman, M.
1986b, \apj, 300, 500 

\reference{CR} Chiang, J. and Romani, R.W. 1994,
\apj, 436, 754 

\reference{PASP} Eikenberry, S.S., Fazio, G.G. and Ransom,
S.M. 1996, PASP, in press

\reference{Crab2} Eikenberry, S.S., Fazio, G.G., Ransom, S.M., Middleditch, J., Kristian, J., Pennypacker, C.R., 1996, in preparation

\reference{HST} Percival, J.W., Biggs,
J.D., Dolan, J.F., Robinson, E.L., Taylor, M.J., Bless, R.C., Elliot,
J.L., Nelson, M.J., Ramseyer, T.F., van Citters, G.W. 1993, \apj, 407,
276 

\reference{RISC} Petroff, M.D., Stapelbroek, M.G., and Kleinhans, W.A.
1987, Appl. Phys. Lett., 51, 406 

\reference{Pravdo} Pravdo, S.H. and Serlemitsos, P.J. 1981, \apj, 246, 484

\reference{SCOTT} Ransom, S.M., Fazio, G.G.,
Eikenberry, S.S., Middleditch, J., Kristian, J.A., Pennypacker, C.R.
and Hays, K.M. 1994, \apjl, 431, L43 

\reference{RY} Romani, R.W. and
Yadigaroglu, I.-A. 1995, \apj, 438, 314 

\reference{SMITH} Smith, F.G., Jones,
D.H.P., Disck, J.S.B. and Pike, C.D. 1988, MNRAS, 233, 305.

\end{references}
\end{document}